# Statistical analysis of the sigma meson considered as two-pion resonance

M. C. Menchaca Maciel and J. R. Morones Ibarra

Facultad De Ciencias Físico-Matemáticas, Universidad Autónoma de Nuevo León,
Ciudad Universitaria, San Nicolás de los Garza, Nuevo León, 66450, México.

It was found that the most simple analysis in the linear σ- model, where the sigma meson couples in vacuum to two virtual pions, predicted the most typical values for the σ–mass and total decay width as $m_\sigma = 486 \pm 7 MeV$ and $\Gamma = 340 \pm 20 MeV$ respectively. It was demonstrate, via statistical analysis, that the experimental values reported by the Fermilab (E791) Collaboration represents approximately 56% of the values predicted by the linear σ- model together with the Breit-Wigner formalism. Additional, it was created the spectral function, in terms of the σ-meson regularized self-energy, which is consistently with the Breit-Wigner distribution and it reproduces reasonably well the total width and observable mass given by the Fermilab (E791) Collaboration and Dalitz plot. The results are a strong evidence that the σ- meson can be considered as two-pion resonance.

Keywords: sigma meson, spectral function, decay width, renormalized mass, Breit-Wigner.

PACS: 14.40; 13.75.L

## 1. INTRODUCTION

There is an increasing interest in study the σ –meson or $f_0(600)$, due to the role it plays in nuclear and hadronic physics [1]. Originally the σ–meson was introduced to fit experimental data and its mass was chosen to reproduce the experimental results. There is a wide variety for defining the mass and width of a particle. Some authors use the pole approach with the mass and width of resonance taken from the position of the pole of the T-matrix [2]. Another way to study the mass and width of resonances is through the use of the spectral function.

Our main motivation for this study was the article written by Nils A. Törnqvist, where the "junk entry" [3] could be interpreted as:

The use of any value in the range from 400MeV to 1200MeV as the σ –meson mass or/and the use of any value in the range from 600MeV to 1000MeV as the σ –meson Breit-Wigner width, without any explanation of how the values have been obtained.

Usually the "junk entry" was used for some theoretically groups in order to fit the spectral function with the experimental values or the σ(PDG) estimation. More recently it was used as "illustration" in the spectral function study [4].

The relativistic Breit-Wigner distribution is usually employed in theoretical and experimental studies to calculate the mass and width of the resonance for the σ –meson [5,6,7], still when some experts say that only the mass and width arising from the pole position should be considered physical quantities.

For these reason in the present work, we will reconsider the linear σ- model and its fundamental studies. The main quality of our study is to consider only the background of the theory and the statistical analysis to test the hypothesis that the relativistic Breit-Wigner distribution does not really give a satisfactory description of the scalar resonance.

In Sec. 2 our purpose will be to construct a modified σ -meson propagator without $m_\sigma$. The mass of the sigma meson only can be find and used after a reliable treatment. In this section, the important outcome is to come to the conclusion that the divergence of the real part of the self-energy is well under control by a different technique.

In Sec. 3 we obtain the values of $m_\sigma$ using a system of equations and restrictions for the behavior of the real part of the inverse propagator. Also, we establish the reason of why the Breit-Wigner distribution can be unsuccessful to describe the resonance. Finally, the full decay width is calculated in Sec. 4. and we conclude that the BW mass and the maximum of the distribution is the same, for the original BW probability density. Also, we demonstrate that there exist a function that satisfies the BW probability density. Therefore, reference [4] must be used only as illustration and not as a real analysis about the resonance of the σ-meson.

## 2. MODIFIED σ -MESON PROPAGATOR AND ANALITICAL EXPRESION FOR THE SELF-ENERGY

*2.1. Problem to solve.*

In the calculation of the scalar one-loop integral, the divergence for the real part of the self-energy is absorbed into the renormalization of the bare mass $m_\sigma^0$, usually as:

$$m_\sigma^2 = \left(m_\sigma^0\right)^2 + \text{Re}\,\Sigma\left(m_\sigma^2\right) \tag{1}$$

$m_\sigma$ represents the observable mass, therefore Eq. (1) implies the knowing of the observable mass or the use of "any entry".

*2.2. Formalism.*

The modified σ -meson propagator is obtained through the modification of the free σ -meson propagator in the one loop approximation. The full propagator can be calculated with the chain approximation method.

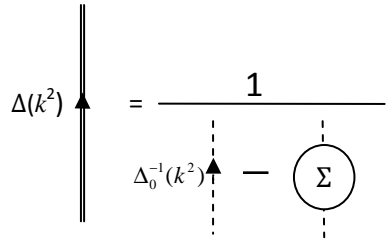

**Figure 1**. Sum over all repetitions of all self-energy parts of any number of pion "bubbles" inserted into the bar σ-meson propagator $\Delta_0(k^2)$.

As indicated in FIG.1, the dressed or modified σ – meson propagator obtained from the Dyson equation is

$$\Delta\left(k^2\right) = \frac{1}{\Delta_0^{-1}(k^2) - \Sigma\left(k^2\right)} \tag{2}$$

$$\Delta_0^{-1}\left(k^2\right) = k^2 - \left(m_\sigma^0\right)^2 \tag{3}$$

$m_\sigma^0$ is the bare mass of the σ -meson and the self-energy $\Sigma\left(k^2\right)$ needs to be calculated.

The simplest renormalizable π-σ coupling obtained from an extension of the σ-model is

$$L_{\sigma\pi\pi}^{eff} = -\lambda \vec{\pi}\cdot\vec{\pi}\sigma \tag{4}$$

where

$$\lambda = \sqrt{\frac{3}{2}}\, g_{\sigma\pi\pi} m_\pi \tag{5}$$

$g_{\sigma\pi\pi}$ is dimensionless and real and becomes from the relation of the σ model between the masses and the coupling constant of σ and the ππ system:

$$g_{\sigma\pi\pi} = \frac{m_\sigma^2 - m_\pi^2}{2f_\pi^2} \tag{6}$$

$f_\pi = 92.4\text{MeV}$ is the pion decay constant and $m_\pi = 139.57\text{MeV}$ the pion mass.

The self-energy has the analytical expression given by the two-pion-bubble integral and the study is in the σ –meson rest frame $k = k^\mu = (k_0, \vec{0})$.

$$-i\Sigma(k^2) = \frac{3g_{\sigma\pi\pi}^2 m_\pi^2}{2} \int \frac{d^4q}{(2\pi)^4} \cdot \frac{1}{q^2 - m_\pi^2} \cdot \frac{1}{(q-k)^2 - m_\pi^2} \tag{7}$$

The constants are suitable for $\vec{\pi} \cdot \vec{\pi}\sigma$ coupling of charge-symmetric pions to the σ meson [8].
After integration, the imaginary part of the self energy is

$$\text{Im}\Sigma(k^2) = -\frac{3}{32\pi} g_{\sigma\pi\pi}^2 m_\pi^2 \sqrt{1 - \frac{4m_\pi^2}{k^2}} \tag{8}$$

The regularized principal part of the self-energy was obtained via the twice-subtracted dispersion relation and its analytical expression is:

$$\text{Re}\Sigma^R(k^2) = \frac{g_{\sigma\pi\pi}^2}{64\pi^2 \sqrt{k^2(k^2 - 4m_\pi^2)}} [(k^2 - 12m_\pi^2)\sqrt{k^2(k^2 - 4m_\pi^2)} + 6m_\pi^2(k^2 - 4m_\pi^2) Ln\left|\frac{\sqrt{k^2(k^2 - 4m_\pi^2)} + k^2}{2m_\pi^2} - 1\right|] \tag{9}$$

The renormalized self-energy is constructed from Eq. (8) and Eq.(9)

$$\Sigma^R(k^2) = \text{Re}\Sigma^R(k^2) + i\,\text{Im}\Sigma(k^2) \tag{10}$$

*2.3. Solution.*
We introduce the concept of the renormalized mass as:

$$m_R^2 = (m_\sigma^0)^2 + \text{Re}\Sigma_0(k^2) \tag{11}$$

$\text{Re}\Sigma_0(k^2)$ is an infinite quantity that cancel the infinite terms of $\text{Re}\Sigma(k^2)$ and $m_R$ is a new parameter.
The modified σ – meson propagator becomes

$$\Delta(k^2, m_R^2) = \frac{1}{k^2 - m_R^2 - \text{Re}\Sigma^R(k^2) - i\,\text{Im}\Sigma(k^2)} \tag{12}$$

## 3. CALCULATION OF THE RENORMALIZED AND OBSERVABLE MASS
We start with the basic concepts of the Breit-Wigner formalism. We demonstrate that imposing correct criteria rejects that we can use any value between 400-1200 MeV as the sigma mass.

*3.1 Breit-Wigner mass*
The observable mass of σ is defined as a zero of the real part of the inverse propagator :

$$\text{Re}\left[\Delta^{-1}(k^2)\right] = 0 \tag{13}$$

Any value of $k$ that satisfies Eq.(13) must be the peak of the resonance or the BW mass and this implies that $k^2 = k^2_{max} = m^2_\sigma$. We note that Eq. (13) can be solve independently of $g_{\sigma\pi\pi}$, since Eq.(6) can be rewritten as

$$g_{\sigma\pi\pi} = \frac{k^2 - m^2_\pi}{2f^2_\pi}\bigg]_{k^2=m^2_\sigma} \qquad (14)$$

therefore, if the sigma meson exists as two-pion resonance, the σ –mass is restricted to satisfy, at the same time, Eq.(13) and Eq.(14).
This allow us to change $g_{\sigma\pi\pi}$ in Eq.(8) and Eq.(9) in terms of $k$ in order to find all ordered pairs $(k, m_R)$ such that:

$$\text{Re}\left[\Delta^{-1}(k^2, m^2_R)\right] = 0 \qquad (15)$$

In our scheme, $m_R \geq 0$ and $m_\sigma \geq 2m_\pi$, the results obtained with our formulas are $0 \leq m_R \leq 423.553$ MeV and $2m_\pi < m_\sigma \leq 649.14$ MeV. As we expected Eq.(15) is described by a simple closed curve when $k$ and $m_R$ are not restricted. In addition, the curve has the form of a simple Lissajous figure or Bowditch curve (See Fig.2).

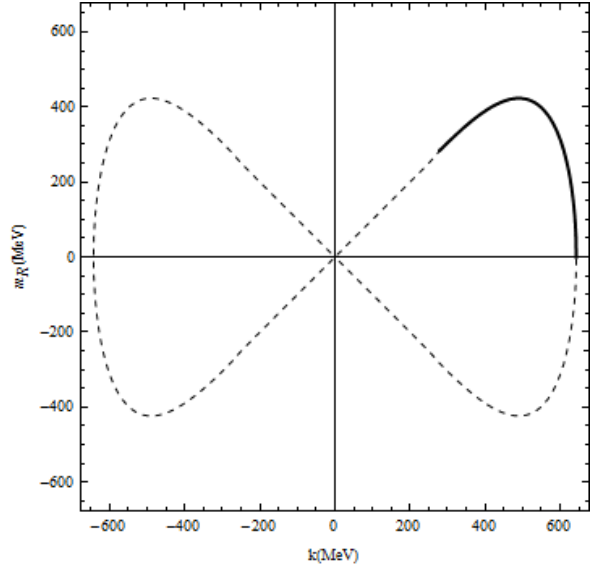

**Figure 2**. Contour plot of Eq.(15). The solid curve represent the possible solutions when $m_R \geq 0$ and $m_\sigma \geq 2m_\pi$. The dashed curve shown the solutions with $(k, m_R)$ as free parameters (without restrictions).

When $m_R \geq 0$ and $m_\sigma \geq 2m_\pi$, Eq.(15) can be described by the parametric equation

$$m_\sigma = 642.980 \, \text{Sin}[t + 0.08], \quad m_R = 423.554 \, \text{Sin}[2t + 0.0004], \quad 0.37 \leq t \leq 1.5 \qquad (16)$$

The importance of this result is because at resonance, the oscilloscope displays a Lissajous figure. Eq.(16) remarks the value about the use of Eq.(11) instead Eq.(1) given in section 2.
At this point, we need a brief discussion about the implications when Eq.(15) is used to obtain the BW mass. First we have that

$$\text{Re}\left[\Delta^{-1}(k^2, m^2_R)\right] = \text{Re}[k^2 - m^2_R - \text{Re}\Sigma^R(k^2) - i\,\text{Im}\Sigma(k^2)] \qquad (17)$$

Eq.(15) implies

$$k^2 = m^2_\sigma = m^2_R + \text{Re}\Sigma^R(m^2_\sigma) \qquad (18)$$

Eq.(12) becomes

$$\Delta(k^2) = \frac{1}{k^2 - m_\sigma^2 - i\,\text{Im}\,\Sigma(k^2)} \tag{19}$$

Eq.(19) has the form of the propagator to second order for an unstable particle. Now we are sure about the validly of Eqs.(4)-(12) and we can continued with the Breit-Wigner formalism.

*3.2 Behavior for the real part of the inverse propagator*

Eq.(17) must be not differentiable at $m_\sigma = 2m_\pi$ and its solution is unique. Not all ordered pairs $(k, m_R)$ satisfy the new conditions. Eq.(17) assumes different shapes for different values of the mass (See Figure 3).

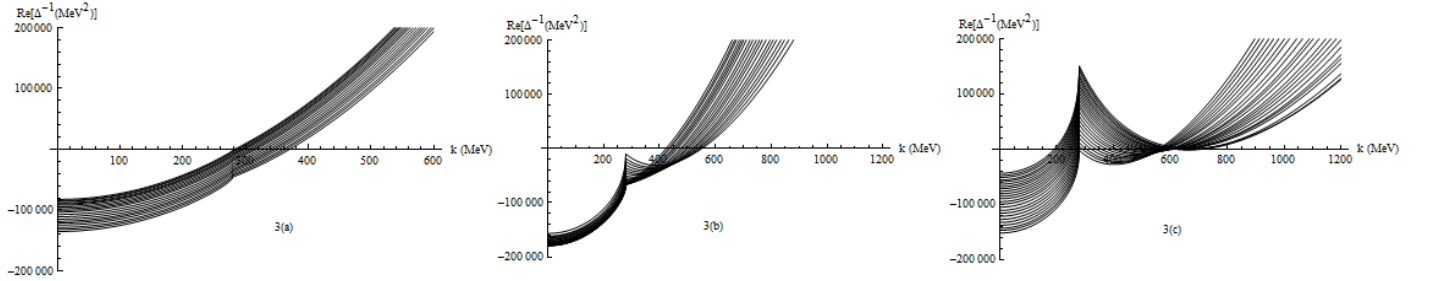

**Figure 3**. Behavior of real part of the inverse propagator. 3(a) for $2m_\pi < m_\sigma < 417$ MeV the graph has a distorted shape for a BW resonance. 3(b) for $417 \leq m_\sigma \leq 555$ MeV the graph has the shape for a correct BW resonance. 3(c) for $555 < m_\sigma \leq 649.14$ MeV the graph cannot represent a BW resonance.

Using all the conditions the mass is restricted to be $417 \leq m_\sigma \leq 555$ MeV, the minimum value that the physical mass can take is $m_\sigma = 417$ MeV, this agrees with quark model that predict that the mass cannot be as low as 390 MeV [9].

$$m_\sigma = 642.980\,\text{Sin}[t + 0.08], \quad m_R = 423.554\,\text{Sin}[2t + 0.0004], \quad 0.62 \leq t \leq 0.96 \tag{20}$$

*3.3 Statistical analysis for the $\sigma$–mass*

**Table 1.** $\sigma$–mass Descriptives

| $m_\sigma$ (MeV) | | Statistic | Std. Error |
|---|---|---|---|
| Mean | | 486.00 | 3.416 |
| 95% Confidence Interval for Mean | Lower Bound | 479.25 | |
| | Upper Bound | 492.75 | |
| 5% Trimmed Mean | | 486.00 | |
| Median | | 486.00 | |
| Variance | | 1.622E3 | |
| Std. Deviation | | 40.270 | |
| Minimum | | 417 | |

| | | |
|---|---:|---:|
| Maximum | 555 | |
| Range | 138 | |
| Interquartile Range | 70 | |
| Skewness | 0.000 | 0.206 |
| Kurtosis | -1.200 | 0.408 |

Theoretically, such well-behaved values should have the same mean, 5% trimmed mean and median. Table 1. allows to say that the mean describe the most typical value in the set of solutions that we found for the physical mass. The skewness statistic is 0.00 this means that the sample is symmetric. The Kolmogorov-Smirnov Test does not reject the hypothesis of normality (See Table 2) and the z-score of the data does not differ from a normally distributed variable with mean 0 and variance 1 (See figure 4). Therefore, we can use the T Test to test whether the mean of $m_\sigma$ differs from the central values given by BES, FermiLab, Dalitz plot and the Roy equation.

**Table 2.** Test of Normality for the observable σ –mass

| | Kolmogorov-Smirnov[a] | | |
|---|---:|---:|---:|
| | Statistic | df | Sig. |
| Observable mass | 0.060 | 139 | 0.200[*] |

a. Lilliefors Significance Correction

*. This is a lower bound of the true significance.

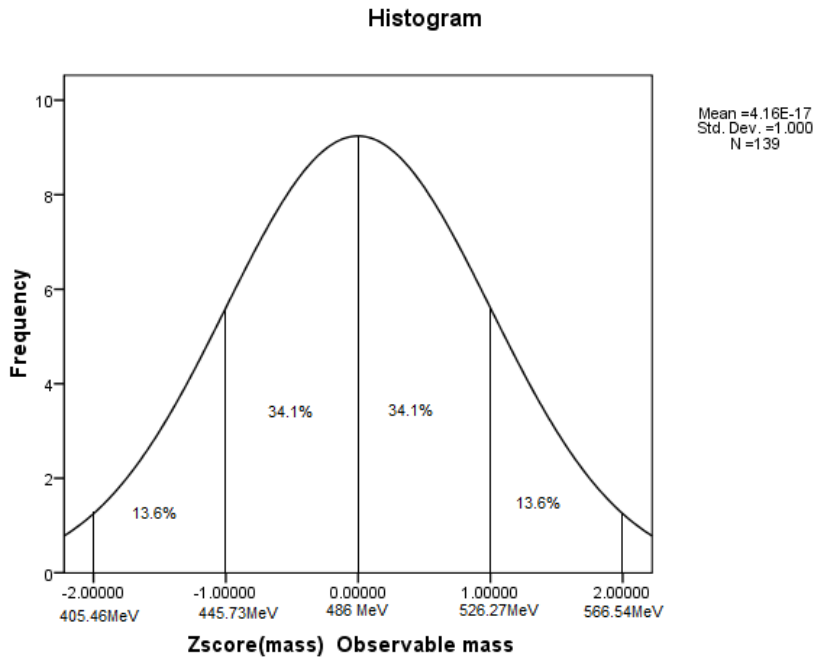

**Figure 4.** z-scores and Normal Probabilities for the physical mass.

**Table 3.** T-Test for each central value

| GROUP | Test value | t | df | Sig. (2-tailed) | Mean Difference | 95% Confidence Interval of the Difference | |
|---|---|---|---|---|---|---|---|
| | | | | | | Lower | Upper |
| BES | 390 | 28.106 | 138 | 0.000 | 96.000 | 89.25 | 102.75 |
| Fermilab | 483 | 0.878 | 138 | 0.381 | 3.000 | -3.75 | 9.75 |
| Dalitz Plot | 478 | 2.342 | 138 | 0.021 | 8.000 | 1.25 | 14.75 |
| Roy Equation | 441 | 13.175 | 138 | 0.000 | 45.000 | 38.25 | 51.75 |

In the T-Test the p value (Sig. (2-tailed) =0.000) is less than 0.05 for the BES Collaboration and Roy Equation, indicating that the average (486MeV) is significantly larger than 390MeV and 441MeV.

For the Fermilab experiment the p values (Sig. (2-tailed) =0.381) is greater than 0.05 therefore the null hypothesis that the average of $m_\sigma$ equals 483 MeV is not rejected. The difference for Dalitz plot and the average of $m_\sigma$ is not rejected at the 0.01 level (Sig=0.021). These results agree with the normal probabilities for the physical mass (See Table 4 and figure 4).

**Table 4.** Normal Probability in the given interval.

| Group | $m_\sigma (MeV)$ | Normal Probabilities |
|---|---|---|
| BES | $390^{+60}_{-36}$ | 18.6 % |
| Fermilab | $483 \pm 31$ | 55.8 % |
| Dalitz Plot | $478^{+24}_{-23} \pm 17$ | 42.4% |
| Roy Equation | $441^{+16}_{-8}$ | 14.2% |

Table 4 says to us, that the probability to obtain the values reported by the Fermilab is 55.8%. Therefore, if the σ – meson exist, the most typical values predicted by the sigma model in concert with the BW mass have been discovered by the Fermilab (E791) Collaboration.

## 4. CALCULATION OF THE TOTAL DECAY WIDTH

In 1976 Earle L. Lomon [10] established that $Im\Sigma(k^2) \approx -M_\sigma \Gamma_\sigma$ when $k^2$ is near $M_\sigma^2$

$$\Gamma_\sigma \equiv -\frac{Im\Sigma\left(m_\sigma^2\right)}{m_\sigma} \qquad (21)$$

This definition arise from the propagator of the unstable particle and was related with the relativistic Breit-Wigner structure of the resonance. Perhaps Eq. (21) is the most important result, because many authors take this relation regardless of how it has been calculated [11,12,13]. In reference [11] the authors used $m_\sigma$=385.4MeV and $g_{\sigma\pi\pi}$=1.6MeV to find the decay width ($\Gamma_\sigma$ = 173MeV), they found that the decay width of the sigma meson was smaller than those reported by the Fermilab(E791) and BES Collaborations.

Far away of the new calculations given in the references [11-13], the imaginary part that Lomon used in his calculations was:

$$\mathrm{Im}\Sigma\left(M_\sigma^2\right) = -\frac{3}{8\pi} g_{\sigma\pi}^2 \sqrt{1 - \frac{4m_\pi^2}{M_\sigma^2}} \tag{22}$$

Using the Lomon's equations and the values in reference [11], the decay width is $\Gamma_\sigma = 547\text{MeV}$, the decay width of the sigma meson is larger than those reported by the Fermilab(E791) and BES Collaborations. If the values given by the Fermilab(E791) or BES experiment are right, we have that the earliest calculations are overestimated and the new calculations are underestimated for the total decay width. Eq.(21) is unsuccessful to describe the total decay width of σ. (For a complete discussion see [14])

*4.1 Total decay width from the Breit-Wigner formalism.*
Using the fact that $\Sigma(k^2)$ is at least second order in coupling constant (See figure 1):

$$\Delta\left(k^2\right)^{(2)} = \frac{1}{k^2 - (m_\sigma - \frac{1}{2}i\Gamma)^2} \tag{23}$$

This implies that $\Delta$ has a pole near $k^2 = (m_\sigma - \frac{1}{2}i\Gamma)^2$, this pole has no direct physical significance because is in the analytic continuation of $\Delta$ across the cut. When we perturb the meson field then $\Gamma$ is the total width at half the maximum of the probability density:

$$\frac{dP}{d^4k} \propto \frac{m_\sigma \Gamma}{\left[k^2 - m_\sigma^2\right]^2 + (m_\sigma \Gamma)^2} \tag{24}$$

Since, $k = (k_0, \vec{0}) = (E, \vec{0})$ and focus on $E \simeq m_\sigma \Rightarrow k^2 - m_\sigma^2 = 2m_\sigma(E - m_\sigma)$, the probability density reduce to:

$$SBW(k) = \frac{\phi\left(\frac{\Gamma}{2}\right)^2}{[k - m_\sigma]^2 + \left(\frac{\Gamma}{2}\right)^2} \tag{25}$$

The constant φ normalize the integral of SBW in terms of k. We found that the spectral function that satisfies Eq. (25) in terms of the imaginary and real part of the self-energy is given by the function M(k):

$$M(k) = \frac{\phi_M \left[\mathrm{Im}\Sigma\left(k^2\right)\right]^2}{2k^2 \left[k - \sqrt{m_R^2 + \mathrm{Re}\Sigma^R\left(k^2\right)}\right]^2 + \left[\mathrm{Im}\Sigma\left(k^2\right)\right]^2} \tag{26}$$

The full width has the relation

$$\Gamma \approx 2.662\Gamma_\sigma - 78.029 MeV = -\frac{2.662\,\mathrm{Im}\Sigma\left(m_\sigma^2\right)}{m_\sigma} - 78.029 MeV \tag{27}$$

The standard error for the constant is 0.707 and 0.004 for the coefficient.

| Parameters | Dalitz Plot | Fermilab |
|---|---|---|
| $m_\sigma$ (MeV) | 478 | 483 |
| $\Gamma$ (MeV) | 324±21 | 338±24 |
| $\Gamma_\sigma$ (MeV) | 147.9019 | 153.9925 |
| M(k) total width (MeV) | 314.0940 | 330.4690 |
| Predicted Value (MeV) | 315.72999 | 331.94503 |
| Adjusted Predicted Value (MeV) | 315.74283 | 331.95602 |

**Table 5** Comparison of widths

In Table 5 the first two rows give the mean values reported by each group. The third row gives the value predicted using Eq.(21). Fourth row gives the value at half the maximum of the M(k) function. The predicted value is calculated via Eq.(27) and the adjusted predicted value is calculated using the standard error for the constant and coefficient of Eq.(27). As we can see the values predicted M(k) and Eq.(27) are less than the values reported, however all of them are into the 95% Confidence Interval given by each group. Also, is clearly that Eq.(21) is unsuccessful to describe the total decay width.

Now we use the Fermilab(E791) and Dalitz Plot central values of the σ-mass to get the value of the renormalized mass and plot Eq.(26) and Eq.(25) to provide evidence that give the same results.

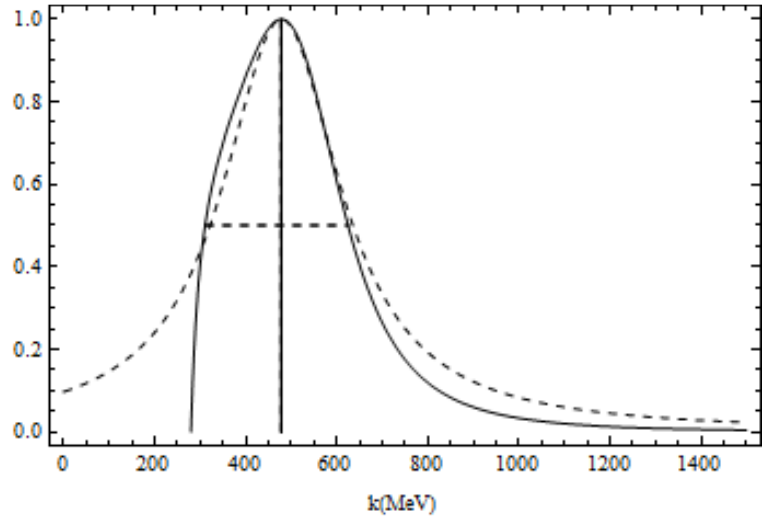

**Figure 5** Function M(k) is the continued curve and SBW(k) is the dashed curve.

In Figure 5 we plot the curves with $\phi_M = \phi = 1$, only for illustration. For both curves the maximum and the observable mass match (478MeV), the coupling constant is about 12.24, and the total decay width is 314.094MeV. The value is into the 95% Confidence Interval given as 324MeV $\pm$21MeV.

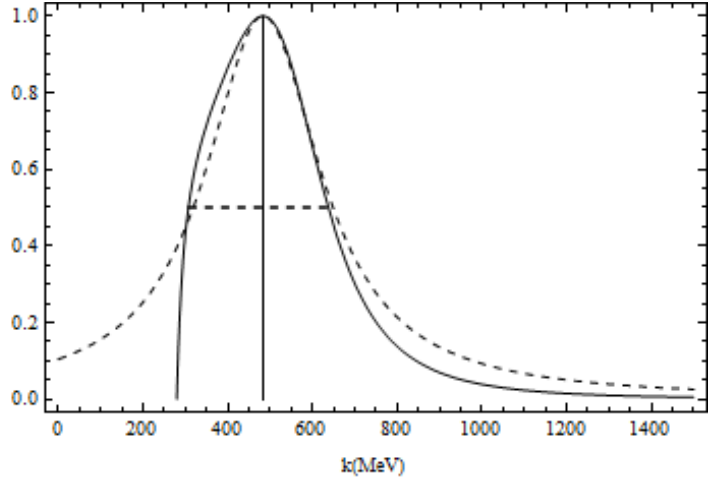

**Figure 6** Function M(k) and SBW(k) for Fermilab central value.

In Figure 6, again $\phi_M = \phi = 1$, the maximum and the observable mass match (483MeV), the coupling constant is about 12.5214, and the total decay width is 330.469MeV, while the value reported was 338MeV. However, our value is into the 95% Confidence Interval given as 338±24.

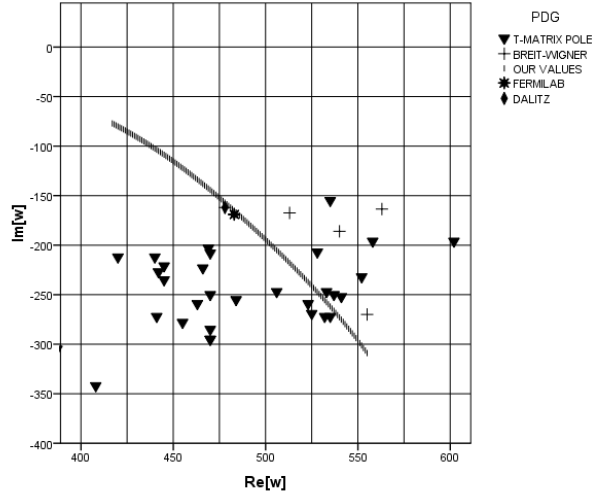

**Figure 7.** Positions of the σ resonance given by the function M(k) and some pole positions as listed in the PDG.

Figure 7 and Figure 8 compare our results for $417 \leq m_\sigma \leq 555$ MeV and the pole position listed by the PDG [15]. + represent the mass and width parameter as $m_\sigma - \frac{1}{2}i\Gamma$, the star represent the E791 experiment and $w = s^{1/2}$ represent the pole.

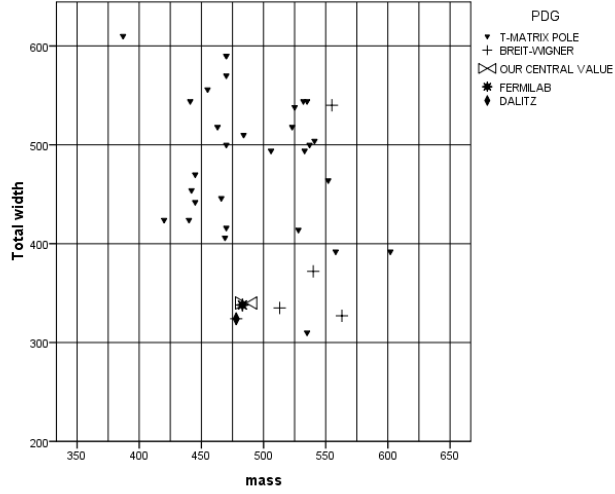

**Figure 8.** Mean value of the mass and its total width for the function M(k) vs. others results.

CONCLUSIONS

The Fermilab (E791) Collaboration [16] in their observation of the $D^+ \to \sigma\pi^+ \to \pi\pi\pi$ decay found the experimental values $m_\sigma = 483 \pm 31 MeV$ and $\Gamma = 338 \pm 48 MeV$. We found that the most typical values are $m_\sigma = 486 \pm 7 MeV$ and the correspondent total width for the central values $\Gamma = 340 \pm 20 MeV$, from the statistical analysis we have that these values are most likely to be the result of the mass and width for the sigma meson considered as two-pion resonance and the difference is not even close to being statistically significant for the mass and width given by the Fermilab experiment and Dalitz plot.

The nature of the $f_0(600)$ or sigma meson is far from being resolved. The recent experimental and theoretically result supports the existence of σ-pole with mass 440-540MeV and width 250-540 MeV [17]. The Breit-Wigner distribution predicts that the mass of σ is between 417-555MeV and its width 150-618MeV. We conclude that the values reported by Fermilab (E791) are more likely for the process $\sigma \to \pi\pi$, if the σ-meson exist. Also, the BW formalism describes the experiment.

ACKNOWLEDGEMENTS
The authors gratefully acknowledge PhD Scholarship Programmer from CONACYT
.